\documentclass[12pt]{article}

\usepackage[totalheight = 23.5cm, totalwidth = 17.8cm]{geometry}
\usepackage{amssymb,amsmath,amsfonts,amsbsy,graphicx,bm}

\usepackage{color,cancel,ulem, hyperref}

\newcommand{\dis}[1]{\begin{equation}\begin{split}#1\end{split}\end{equation}}

\begin{document}

\begin{titlepage}

\begin{center}

 {\setlength{\baselineskip}%
{1.5\baselineskip}
{\LARGE \bf 
 Reheating surface volume renormalization and eternal inflation under broken scale invariance 
}
\par}

\vskip 1.0cm

{\large
Min-Seok Seo$^{a}$ 
\footnote[0]{e-mail : minseokseo57@gmail.com}
}

\vskip 0.5cm

{\it
$^{a}$Department of Physics Education, Korea National University of Education,
\\ 
Cheongju 28173, Republic of Korea
}

\vskip 1.2cm

\end{center}

\begin{abstract}
Recently, Cotler, Ivo, and Maldacena have shown that under scale invariance,  eternal inflation is interpreted as a failure of the reheating surface volume renormalization due to an ill-defined scaling dimension. 
Meanwhile, observations suggest that scale invariance is slightly broken, which is parametrized by deviation of the spectral index from unity, $\varepsilon=n_s-1$.
Motivated by this, we study how the scale invariance breaking effect modifies the no-eternal inflation condition as the renormalizability of the reheating surface volume.
For this purpose, we adopt two independent approaches: investigating the behavior of the normal ordered reheating surface volume operator under a shift of the curvature perturbation, and analyzing the coarse-graining structure to obtain the scaling dimension. 
Both approaches commonly reveal that as long as $\varepsilon$ is negative, which is consistent with observations, the fractal structure of the reheating surface volume breaks down at macroscopic scale, moderating the no-eternal inflation condition.

\end{abstract}

\end{titlepage}

\newpage

\section{Introduction}

The idea of eternal inflation \cite{Steinhardt:1982kg, Vilenkin:1983xq, Linde:1986fc, Linde:1986fd} (for reviews, see, e.g., \cite{Goncharov:1987ir, Guth:2007ng}) exhibits the remarkable effect of large quantum fluctuations on the macroscopic structure of the early universe.
This is realized by two ingredients: an approximate scale invariance and a large amplitude of curvature perturbation.
More concretely, when  curvature perturbation crosses the horizon, it behaves like the classical fluctuation  of time that evolves stochastically and accumulates  over time \cite{Starobinsky:1986fx, Starobinsky:1994bd} (for a review, see, e.g., \cite{Cruces:2022imf}).
When the fluctuation amplitude is sufficiently large, such a stochastic evolution of the fluctuations results in the strong wrinkling of the reheating surface.
Furthermore, if the scale invariance is an exact symmetry, the reheating surface takes on the complete self-similar fractal pattern \cite{Aryal:1987vn, Winitzki:2001np, Winitzki:2008ph}, such that the large wrinkling at the microscopic scale recurs at the macroscopic scale as well.
Eventually, inflating regions are eternally self-produced,  leading to the infinite reheating surface volume.

Whether eternal inflation can take place depends on how much the background geometry during inflation deviates from de Sitter (dS) space, which is encoded in the slow-roll parameters.
First, the fluctuation amplitude becomes small enough to prevent eternal inflation provided that $\epsilon \equiv -{\dot H}/H^2$ is larger than $[3/(4\pi^2)](H^2/M_{\rm Pl}^2)$, where $H$ is the Hubble parameter and $M_{\rm Pl}^2=(8\pi G)^{-1}$ \cite{Creminelli:2008es, Dubovsky:2008rf, Dubovsky:2011uy}. 
 Indeed, several studies have suggested that this condition must be satisfied for consistency with quantum gravity \cite{Matsui:2018bsy, Dimopoulos:2018upl, Kinney:2018kew, Brahma:2019iyy, Rudelius:2019cfh, Wang:2019eym, Seo:2021bpb, Seo:2022uaz} (see also \cite{Arkani-Hamed:2007ryv}), motivated by the difficulty and the subtlety of string model building \cite{Danielsson:2018ztv, Obied:2018sgi} (see also \cite{Andriot:2018wzk, Garg:2018reu, Cicoli:2018kdo, Ooguri:2018wrx, Andriot:2018mav} for a refinement).
Moreover, a slight breaking of scale invariance, as encoded in the spectral index and observed in cosmic microwave background \cite{Planck:2018jri}, breaks the fractal pattern, moderating the divergence of the reheating surface volume \cite{Dubovsky:2008rf, Lewandowski:2013aka}. 
 To reach this conclusion, \cite{Dubovsky:2008rf} qualitatively illustrated the moderation under the adiabatic approximation by  introducing a reflecting barrier that restricts the inflaton field excursion.
 A more detailed treatment was provided in \cite{Lewandowski:2013aka}, which demonstrated the moderation numerically by considering the chaotic inflation scenario.

 The stochastic behavior of the fluctuations can be analytically described by solving the Fokker-Planck equation.
 On the other hand, a recent study in  \cite{Cotler:2026lna} (see also \cite{Cotler:2026flq}) has pointed out  that  this behavior also can be understood in the context of the renormalization of the reheating surface volume as a quantum operator, emphasizing the structural equivalence between Liouville theory and stochastic inflation.
 In more detail, the authors of \cite{Cotler:2026lna} considered the   scaling dimension of an operator of the same form as the reheating surface volume, which is obtained by coarse-graining microscopic fluctuations.
 In the presence of scale invariance, it is given by the Knizhnik-Polyakov-Zamolodchikov (KPZ) formula \cite{Polyakov:1981rd, Knizhnik:1988ak}, which states that if the amplitude of the fluctuation is not sufficiently small ($\epsilon < [3/(4\pi^2)](H^2/M_{\rm Pl}^2)$ in the language of inflationary cosmology),  the coarse-grained volume diverges, leading to  a failure of  the renormalization.
 This is also reflected in the ill-defined scaling dimension.
   
 Meanwhile, the fact that the power spectrum of   curvature perturbation slightly breaks  scale invariance invites us  to explore how such a breaking  is  described in this   renormalization approach, which is the aim of the current work.  
Our discussion is organized as follows. 
 In Section \ref{Sec:Rev}, we review how the breaking of scale invariance  in the power spectrum of  curvature perturbation is parametrized, and establish the notation.
 Since the parametrization is a kinematical description, it does not depend on the specific model of inflation such as the form of the potential.
 To illustrate the effects of scale invariance breaking in a simple and clear way, we only consider the leading effect parametrized by the spectral index, while a more accurate description is possible by including higher order effects as well. 
  Based on this, in Section \ref{Sec:Violation}, we obtain the correction to the scaling dimension of the reheating surface volume operator using the renormalization approach of \cite{Cotler:2026lna}.
  Instead of  explicitly calculating probability distribution for fluctuations using the Fokker-Planck equation, we take two independent approaches.
   First, in Section \ref{subsec:appr1}, we identify the shift of  curvature perturbation with the renormalization parameter by fixing the physical distance cutoff, from which the scaling dimension of the reheating surface volume in the form of the normal ordered operator is obtained. 
  Meanwhile, in Section  \ref{subsec:appr2},  we obtain the same result from the structure of coarse-graining by separating the last step and investigating the behavior of the remaining steps.
  Since it relies on the setup and coarse-graining considered in \cite{Cotler:2026lna}, Section  \ref{subsec:appr2} begins with a brief review of this. 
  While the probability distribution used in \cite{Cotler:2026lna} is obtained by solving the Fokker-Planck equation, it also can be understood by considering features of the first passage process, which are explained in  Appendix \ref{App:ProbDist}.
 Section \ref{Sec:Discussion} is devoted to the interpretation of our results and discussions on their physical implications.
 Then we conclude.

 \section{Power spectrum and scale invariance breaking }
 \label{Sec:Rev}
 
 We begin our discussion by reviewing the structure of the power spectrum and how a slight breaking of scale invariance is parametrized in it. 
 In cosmological perturbation theory, curvature perturbation $\zeta$ corresponds to the trace fluctuation of the spatial metric.
  \footnote{More precisely, since the quasi-dS background during inflation slightly breaks the timelike isometry of dS space, $\zeta$ becomes physical by absorbing the inflaton fluctuation, forming the gauge invariant combination called the Mukhanov-Sasaki variable  \cite{Mukhanov:1985rz, Sasaki:1986hm} (for recent discussions, see, e.g., \cite{Seo:2025tsw, Seo:2026kxs}),
 \dis{{\cal R}=\zeta-\frac{H}{\dot{\phi}}\varphi= \zeta-\frac{1}{\sqrt{2\epsilon}M_{\rm Pl}}\varphi.\nonumber }
 Choosing the comoving gauge $\varphi=0$, ${\cal R}$ is identified with $\zeta$.}
Then the metric of the reheating surface is given by 
 \dis{ds^2=e^{2\zeta} a(t_f)^2 d\mathbf{x}^2 \equiv e^{2\zeta} d \hat{s}^2, \label{eq:metric}}
 where $a(t_f)$ is the scale factor at the end of inflation.
 In momentum space, the two-point function of $\zeta$ is given by \cite{Mukhanov:1981xt, Starobinsky:1982ee, Guth:1982ec, Hawking:1982cz} (for reviews, see, e.g., \cite{Lidsey:1995np, Lyth:1998xn, Riotto:2002yw})
 \dis{\langle \zeta_{\mathbf k}\zeta_{-{\mathbf k}}\rangle=\frac{2\pi^2}{k^3}\frac{1}{2 \epsilon M_{\rm Pl}^2}\Big(\frac{H}{2\pi}\Big)^2 \equiv \frac{2\pi^2}{k^3} {\cal P}_{\zeta}(k).}
 If the power spectrum ${\cal P}_{\zeta}(k)$ is independent of $k$, it  is said to be scale invariant, and as will be shown later, $\langle \zeta(\mathbf{x})\zeta(\mathbf{y})\rangle$, the two-point function in position space, is given by the logarithmic function of $|\mathbf{x}-\mathbf{y}|$.
 However, this in general does not happen as both $\epsilon$ and $H$ depend on time.
 Since  the values of  $\epsilon$ and $H$ in ${\cal P}_{\zeta}(k)$ are taken at the moment $t_e$ of horizon exit, satisfying $k= aH \simeq e^{Ht_e}H$, hence $dk/dt_e\simeq Hk$, one can parametrize the breaking of scale invariance by the spectral index $n_s$ defined by \cite{Liddle:1992wi, Stewart:1993bc}  
 \dis{n_s-1=\frac{d \log {\cal P}_\zeta}{d \log k}=-2\epsilon-\tilde{\eta},\label{eq:ns-1}}
 where $\tilde{\eta}=\dot{\epsilon}/(H\epsilon)$ (a dot here denotes the derivative with respect to time).
  Moreover, the running of the spectral index is quantified  by   $\alpha_s=d n_s/d\log k$ \cite{Kosowsky:1995aa}. 
  
 Meanwhile, ${\cal P}_\zeta(k)$  also can be parametrized to encapsulate its scale dependence in a power-law form.
 For this purpose, we introduce a   pivotal scale $k_0$, such that ${\cal P}_\zeta(k)$ is expanded as \cite{Schwarz:2001vv}
 \dis{\log{\cal P}_\zeta(k)&=\log{\cal P}_\zeta(k_0)+\frac{d \log {\cal P}_\zeta}{d \log k}\Big|_{k=k_0}\log\frac{k}{k_0}+\frac12 \frac{d^2 \log {\cal P}_\zeta}{d (\log k)^2}\Big|_{k=k_0}\log^2\frac{k}{k_0}+\cdots 
 \\
 &=\log {\cal P}_\zeta(k_0)+(n_s(k_0)-1)\log\frac{k}{k_0}+\frac12 \alpha_s \log^2\frac{k}{k_0}+\cdots,}
from which  we obtain
 \dis{{\cal P}_\zeta(k)&=  {\cal P}_\zeta(k_0) \Big(\frac{k}{k_0}\Big)^{n_s(k_0)-1+\frac12 a_s(k_0)\log\frac{k}{k_0}+\cdots}  
 \\
 &= \frac{1}{2 \epsilon(k_0) M_{\rm Pl}^2}\Big(\frac{H(k_0)}{2\pi}\Big)^2 \Big(\frac{k}{k_0}\Big)^{n_s(k_0)-1+\frac12 a_s(k_0)\log\frac{k}{k_0}+\cdots}.\label{eq:P1}}
 Since  ${\cal P}_\zeta(k)$ does not depend on the choice of $k_0$,  $d\log{\cal P}_\zeta(k)/d\log k_0=0$ must be satisfied, which is written as
 {\small 
 \dis{&\frac{d\log{\cal P}_\zeta(k)}{d\log k_0}
 \\
 &=\frac{d \log {\cal P}_\zeta(k_0)}{d \log k_0}+\Big(\frac{d n_s(k_0)}{d \log k_0}\log\frac{k}{k_0}-(n_s(k_0)-1)\Big)+\Big(\frac12\frac{d \alpha_s(k_0)}{d \log k_0}\log^2\frac{k}{k_0}-\alpha_s(k_0)\log\frac{k}{k_0}\Big)+\cdots
 \\
 &=\frac{d \log {\cal P}_\zeta(k_0)}{d \log k_0}+\Big(\alpha_s(k_0)\log\frac{k}{k_0}-(n_s(k_0)-1)\Big)+\Big(\frac12\frac{d \alpha_s(k_0)}{d \log k_0}\log^2\frac{k}{k_0}-\alpha_s(k_0)\log\frac{k}{k_0}\Big)+\cdots
 \\
 &=\frac{d \log {\cal P}_\zeta(k_0)}{d \log k_0}-(n_s(k_0)-1) = -2\epsilon(k_0)-\tilde{\eta}(k_0)-(n_s(k_0)-1)=0, }
 }
 reproducing the relation \eqref{eq:ns-1} at $k=k_0$.

 In the following discussion, in order to investigate the effects of broken scale invariance in a simpler and clear way, we restrict our attention to the case where the running effects parametrized by, for example, $\alpha_s$ is negligible. 
Then $\langle \zeta_{\mathbf k}\zeta_{-{\mathbf k}}\rangle$ is regarded as the $d=3$ and $\varepsilon=n_s-1$ case of the two-point function
\dis{G_2(\mathbf{k})=\frac{(2\pi)^d}{k^d \Omega_{d-1}}{\cal P}_\zeta(k_0)\Big(\frac{k}{k_0}\Big)^\varepsilon,} 
where $\Omega_{d-1}=2\pi^{d/2}/\Gamma(d/2)$.
Taking the Fourier transformation this into position space yields
\dis{G_2(\mathbf{x})=\int \frac{d^dk}{(2\pi)^d} e^{i \mathbf{k}\cdot \mathbf{x}}G_2(\mathbf{k})= {\cal P}_\zeta(k_0)\frac{1}{k_0^\varepsilon}\int \frac{dk}{k}k^\varepsilon\Big(\frac{1}{\Omega_{d-1}}\int d\Omega_{d-1} e^{i \mathbf{k}\cdot \mathbf{x}}\Big). }
While the normalized angular integral in the parenthesis is given by $\Gamma(d/2)(2/kr)^{\frac{d}{2}-1}J_{\frac{d}{2}-1}(kr)$ with $r=|\mathbf{x}|$, since it quickly falls off for $k>1/r$, it is a good approximation to  restrict the integration interval to $[k_{\rm min}, 1/r]$, where  $k_{\rm min}$ is an IR cutoff regulating the IR divergence for $\varepsilon<0$.
 In this case, we can set $e^{i \mathbf{k}\cdot \mathbf{x}}\simeq 1$ such that the normalized angular integral can be approximated by $1$, then one obtains
\dis{G_2(\mathbf{x})&\simeq {\cal P}_\zeta(k_0)\frac{1}{\varepsilon }\Big[\frac{1}{(k_0r)^\varepsilon} - \Big(\frac{k_{\rm min}}{k_0} \Big)^\varepsilon\Big]
\\
&= {\cal P}_\zeta(k_0)\log\Big(\frac{1}{k_{\rm min} r}\Big)\Big[1-\frac{\varepsilon}{2}\log\frac{k_0^2 r}{k_{\rm min}}+{\cal O}(\varepsilon^2)\Big].\label{eq:Gx}}
We note that since $k=a H\simeq e^{Ht_e}H$,  $\log(k_0/k_{\rm min})$ is interpreted as  the number of $e$-folds $N_e(k_0)$ at horizon exit for the $k_0$ mode.
Moreover, the first line shows that when $\varepsilon<0$, $G_2(\mathbf{x})$ exhibits the IR divergence, which is already evident from the power counting, thus $k_{\rm min}$ cannot be sent to $0$.

In the following discussion, we adopt the shorthand notation of  \cite{Cotler:2026lna},
\dis{{\cal P}_\zeta(k_0)=  \frac{1}{2 \epsilon(k_0) M_{\rm Pl}^2}\Big(\frac{H(k_0)}{2\pi}\Big)^2 \equiv \frac{2}{d Q^2(k_0)}\label{eq:P0}}
such that the two-point function in position space is written as 
\dis{G_2(\mathbf{x})=\frac{2}{d Q^2(k_0)}\log\Big(\frac{1}{k_{\rm min} r}\Big)\Big[1- \frac{\varepsilon}{2}\log\frac{k_0^2 r}{k_{\rm min}}+{\cal O}(\varepsilon^2)\Big], \label{eq:G2}}
where $\varepsilon=n_s-1$ parametrizes the breaking of scale invariance.
When scale invariance is exact, i.e., $\varepsilon=0$, $G_2(\mathbf{x})$ is given by the logarithmic function of $r=\mathbf{|x|}$, which is considered in \cite{Cotler:2026lna}.

 \section{Reheating surface volume renormalization under  broken scale invariance}
 \label{Sec:Violation}
 
  We now evaluate the scaling dimension of the reheating surface volume as an operator  to observe how the breaking of scale invariance  parametrized by $\varepsilon$  moderates the condition for the failure of renormalization.
For this purpose, we need to coarse-grain the fluctuations of $\zeta$ between two different distance cutoff scales, following  \cite{Cotler:2026lna}.
As a cutoff, we consider the  smallest length scale measured with respect to the comoving coordinates $\mathbf{x}$, or equivalently,  the fiducial metric $d\hat{s}^2$ in \eqref{eq:metric}.
 This will be called the fiducial distance cutoff (or simply the fiducial cutoff) and denoted by a hat, e.g., $\hat{\epsilon}$.
  
  As can be inferred from the metric \eqref{eq:metric}, the physical distance cutoff $\epsilon$ and the fiducial distance cutoff  $\hat{\epsilon}'$ are related by $\epsilon =e^\zeta \hat{\epsilon}'$. 
 Then even though  $\epsilon$ is given by the fixed value, $\hat{\epsilon}'$ fluctuates along $\zeta$, such that the reheating surface implemented by a constant physical coordinate condition becomes wrinkled in the comoving coordinates.
In the presence of scale invariance, the wrinkling  appears in the self-similar way at every scale, leading to the  fractal structure of the reheating surface.
Furthermore, when the fluctuation of $\zeta$ is  large enough to allow eternal inflation,  the reheating surface becomes strongly wrinkled.
In this case, under the renormalization running from the strongly fluctuating fiducial cutoff  $\hat{\epsilon}'$ to the fixed fiducial cutoff $\hat{\epsilon}$, the reheating surface volume diverges, resulting in the failure of renormalization.
The condition for renormalization failure under an exact scale invariance has been studied in \cite{Cotler:2026lna}, which coincides with the condition that eternal inflation takes place.

Meanwhile, the correction from nonzero $\varepsilon$ breaks the scale invariance, hence the fractal pattern of the reheating surface.
In particular, since high-$k$ modes cross the horizon at a later time, the negative $\varepsilon =d\log{\cal P}_\zeta/d\log k$ implies that  the fluctuation amplitude gets smaller as it evolves into more macroscopic scale.
Then we expect that the condition  for renormalization failure would be relaxed. 
To reach this result, we take two approaches. 
In Section \ref{subsec:appr1}, we consider the normal ordered operator that can be interpreted as the reheating surface volume operator, and observe its behavior under the shift of  $\zeta$.
On the other hand, in Section \ref{subsec:appr2}, the approach similar to \cite{Cotler:2026lna} (explained in the previous paragraph) is taken, and the scaling dimension is obtained by separating the last step of renormalization running from the rest.

\subsection{Approach 1 }
 \label{subsec:appr1}

Let us consider an operator $e^{dh\zeta}$ with  scaling dimension $\Delta = dh$, which can be read off from the behavior of the operator  under $\zeta \to \zeta-\lambda$,
\dis{e^{dh\zeta} \longrightarrow e^{-d h\lambda}e^{dh\zeta} \equiv e^{-\Delta\lambda }e^{dh\zeta}.}
Given the metric \eqref{eq:metric}, the physical reheating surface volume reads $e^{d\zeta}$ (where the constant $a(t_f)$ is absorbed into the redefinition of the spatial coordinates), corresponding to the $h=1$ case of it.
 In order to factor out the divergence captured in the vacuum expectation value, we introduce a normal ordered operator $:e^{\alpha\zeta}:$  having the same scaling dimension as $e^{ d h\zeta}$, following \cite{Cotler:2026lna}.
 More concretely, assuming a Gaussian distribution for the fluctuation of $\zeta$ with variance $\langle\zeta^2\rangle$, the vacuum expectation value is given by  
\dis{\langle e^{\alpha \zeta}\rangle=\int d\zeta e^{\alpha\zeta} \frac{1}{\sqrt{2\pi\langle\zeta^2\rangle}}e^{-\frac{\zeta^2}{2\langle\zeta^2\rangle}}=e^{\frac{\alpha^2}{2}\langle\zeta^2\rangle}.\label{eq:Gauss}}
%Here  $\langle\zeta^2\rangle$, the variance of $\zeta$, can be obtained by coarse-graining fluctuations from some fundamental UV cutoff scale to the scale where $\zeta$ s observed.  
Then the normal ordered operator is defined by 
\dis{: e^{\alpha \zeta} : =\frac{ e^{\alpha \zeta}}{\langle e^{\alpha \zeta}\rangle} = e^{-\frac{\alpha^2}{2}\langle\zeta^2\rangle}e^{\alpha \zeta}.}

 On the other hand, the scaling dimension of  $:e^{\alpha\zeta}:$ can be  evaluated by considering the running of the operator between two fiducial cutoff scales $\hat{\epsilon}_1$ and $\hat{\epsilon}_2$ with a common fixed physical cutoff scale $\epsilon$.
 To see this, we adopt the following notations:
 \begin{itemize}
 \item We denote the field value $\zeta$ at the fiducial cutoff scale $\hat{\epsilon}$ by $\zeta_{\hat \epsilon}$ (thus the relation $\epsilon=e^{\zeta_{\hat \epsilon}}  \hat{\epsilon}$ is satisfied), while $\zeta_{\hat{\epsilon}_1}$ will be simply written as  $\zeta_1$.
  Then one can impose  $\epsilon=e^{\zeta_1}\hat{\epsilon}_1=e^{\zeta_2} \hat{\epsilon_2}$.
 \item Coarse-graining of the fluctuations from $\hat{\epsilon}_1$ to $\hat{\epsilon}_2$ is denoted by $\langle~\rangle_{[12]}$.
 Meanwhile, we represent the variance of $\zeta$ with the cutoff   $\hat{\epsilon}_1$ as $\langle \zeta^2_1 \rangle$. 
\end{itemize}  
Then we can confirm that the operator  $:e^{\alpha\zeta}:$  behaves well under the coarse-graining:  for $\zeta_1=\zeta_2+\delta\zeta$, one finds 
\dis{\langle :e^{\alpha\zeta_1}:\rangle_{[12]}&=e^{-\frac{\alpha^2}{2}\langle\zeta_1^2\rangle}\langle e^{\alpha \zeta_1} \rangle_{[12]}=e^{-\frac{\alpha^2}{2}\langle\zeta_1^2\rangle}\langle e^{\alpha (\zeta_2+\delta \zeta)} \rangle_{[12]}
\\
&=e^{-\frac{\alpha^2}{2}\langle\zeta_1^2\rangle} e^{\alpha\zeta_2}\langle e^{\alpha  \delta \zeta } \rangle_{[12]} = e^{-\frac{\alpha^2}{2}\langle\zeta_1^2\rangle} e^{\alpha\zeta_2}  e^{\frac{\alpha^2}{2} \langle\delta \zeta^2 \rangle_{[12]}},  }  
where for the last equality  \eqref{eq:Gauss} is used. 
Using $\langle\zeta_1^2\rangle=\langle(\zeta_2+\delta\zeta)^2\rangle= \langle\zeta_2^2\rangle+\langle\delta\zeta^2\rangle_{[12]}$, it becomes 
\dis{\langle :e^{\alpha\zeta_1}:\rangle_{[12]}&=
e^{-\frac{\alpha^2}{2}\langle\zeta_1^2\rangle} e^{\alpha\zeta_2}  e^{\frac{\alpha^2}{2} \langle\delta \zeta^2 \rangle_{[12]}}
=e^{-\frac{\alpha^2}{2}\langle\zeta_1^2\rangle} e^{\alpha\zeta_2}  e^{\frac{\alpha^2}{2} (\langle\zeta_1^2\rangle-\langle\zeta_2^2\rangle)}
\\
&=e^{-\frac{\alpha^2}{2}\langle\zeta_2^2\rangle}e^{\alpha \zeta_2}=:e^{\alpha\zeta_2}:, \label{eq:NO}}
as expected.

 Now,  the two-point function given by  \eqref{eq:G2} leads to
\dis{\langle\zeta_1^2\rangle&=\frac{2}{d Q^2(k_0)}\log\Big(\frac{1}{k_{\rm min} \hat{\epsilon}_1 }\Big)\Big[1- \frac{\varepsilon}{2}\log\frac{k_0^2 \hat{\epsilon}_1}{k_{\rm min}}+{\cal O}(\varepsilon^2)\Big]
\\
&=\frac{2}{d Q^2(k_0)}\log\Big(\frac{1}{k_{\rm min}  e^{-\zeta_1}\epsilon  }\Big)\Big[1- \frac{\varepsilon}{2}\log\frac{k_0^2  e^{-\zeta_1}\epsilon}{k_{\rm min}}+{\cal O}(\varepsilon^2)\Big].}
Choosing $k_0=\epsilon^{-1}$, it becomes
\dis{\langle\zeta_1^2\rangle=\frac{2}{d Q^2(\epsilon^{-1})} \Big[\zeta_1+\frac{\varepsilon}{2}\zeta_1^2+({\rm constant})+{\cal O}(\varepsilon^2)\Big],}
from which one finds that the scale invariance breaking effect parametrized by $\varepsilon$ depends on $\zeta_1$ as well.
This indicates that the scaling dimension is no longer a global   constant, but given by a local function of $\zeta_1$, thus determined by the infinitesimal shift around $\zeta_1$. 
Under the infinitesimal shift, i.e., $\delta \zeta \ll \zeta_{1,2}$, we obtain
\dis{\langle\delta \zeta^2 \rangle_{[12]}&=\langle \zeta_1^2\rangle-\langle \zeta_2^2\rangle 
\\
&\simeq \frac{d \langle \zeta_1^2\rangle}{d\zeta} \delta\zeta
  =\frac{2}{d Q^2(\epsilon^{-1})} \big(1+\varepsilon \zeta_1\big)\delta \zeta, \label{eq:varzeta}}
then the relation \eqref{eq:NO} can be written as
\dis{:e^{\alpha\zeta_2}:&=\langle :e^{\alpha\zeta_1}:\rangle_{[12]}=
e^{-\frac{\alpha^2}{2}\langle\zeta_1^2\rangle} e^{\alpha\zeta_2}  e^{\frac{\alpha^2}{2} \langle\delta \zeta^2 \rangle_{[12]}}
\\
&=
e^{-\frac{\alpha^2}{2}\langle\zeta_1^2\rangle} e^{\alpha(\zeta_1-\delta \zeta)}  e^{\frac{\alpha^2}{2} \times \frac{2}{d Q^2(\epsilon^{-1})}  (1+\varepsilon \zeta_1 )\delta \zeta }
\\
&=e^{-\big(\alpha-\frac{\alpha^2}{d Q^2(\epsilon^{-1})}(1+\varepsilon\zeta_1)\big)\delta \zeta} e^{-\frac{\alpha^2}{2}\langle\zeta_1^2\rangle} e^{\alpha \zeta_1}
\\
&=e^{-\big(\alpha-\frac{\alpha^2}{d Q^2(\epsilon^{-1})}(1+\varepsilon\zeta_1)\big)\delta \zeta} :e^{\alpha\zeta_1}:.}
 Therefore, the scaling dimension reads
\dis{\Delta =d h =\alpha-\frac{\alpha^2}{d Q^2(\epsilon^{-1})}(1+\varepsilon\zeta).\label{eq:Delta1}}
In \cite{Cotler:2026lna}, a new quantity $\alpha_h$ defined by $\alpha= d Q(\epsilon^{-1}) \alpha_h$ is introduced, which in our case is given by 
\dis{\alpha_h=\frac{1}{2(1+\varepsilon\zeta)}\big[Q-\sqrt{Q^2-4h(1+\varepsilon\zeta)}\big],\label{eq:ah1}}
where we briefly denote $Q(\epsilon^{-1})$ by $Q$.

 \subsection{Approach 2 }
 \label{subsec:appr2}
 
\subsubsection{Renormalization procedure for reheating surface volume} 
 
 We can also arrive at \eqref{eq:ah1} by taking a different approach,   considering the renormalization of the reheating surface volume more explicitly.
 This was extensively discussed in \cite{Cotler:2026lna} for the case of exact scale invariance, which will be briefly sketched below.  
 Before proceeding, we first summarize the renormalization scales relevant to our discussion, using the notations of \cite{Cotler:2026lna}. 
 The physical and fiducial UV cutoffs are related by
  \dis{\epsilon_{\rm phys} \equiv e^L = e^{\zeta_L}\hat{\epsilon}_{\rm UV} \equiv e^{\zeta_L}e^{-\tau_{\rm UV}}, \quad\quad {\rm hence}\quad\quad   L=\zeta_L-\tau_{\rm UV}.}
 As discussed in the beginning of this section, $\epsilon_{\rm phys}$ (hence $L$) is given by the fixed scale, but $\hat{\epsilon}_{\rm UV}$ (hence $\tau_{\rm UV}$) fluctuates as $\zeta_L$ does. 
Meanwhile, we renormalize the reheating surface volume operator down to the IR cutoff, 
\dis{\epsilon_{\rm IR}= e^{\zeta_{\rm IR}}\hat{\epsilon}_{\rm IR} \equiv e^{\zeta_{\rm IR}}e^{-\tau_{\rm IR}},}
where $\hat{\epsilon}_{\rm IR}$ (hence $\tau_{\rm IR}$) is fixed.
Then  the variations of $\zeta$ and $\tau$ during coarse graining are given by
\footnote{While $\tau$ in \cite{Cotler:2026lna}  corresponds to $t_{\rm tot}$ here, we distinguish the scale (denoted by $\tau$) from the interval between scales (denoted by $t$) by using different notations.}
\dis{\zeta_{t, {\rm tot}}=\zeta_L-\zeta_{\rm IR},\quad\quad t_{\rm tot}=\tau_{\rm UV}-\tau_{\rm IR},\label{eq:param}}
respectively.
We also define $N_0$ as
\dis{e^{N_0}=\frac{\epsilon_{\rm IR}}{\epsilon_{\rm phys}}=\frac{e^{\zeta_{\rm IR}}e^{-\tau_{\rm IR}}}{e^L}=\frac{e^{\zeta_{\rm IR}}e^{-\tau_{\rm IR}}}{e^{\zeta_L}e^{-\tau_{\rm UV}}},}
such that
\dis{N_0&=\zeta_{\rm IR}-\tau_{\rm IR}-L = (\zeta_{\rm IR}-\zeta_L)+(\tau_{\rm UV}-\tau_{\rm IR})
\\
&=-\zeta_{t, {\rm tot}} +t_{\rm tot}\label{eq:N0}}
is satisfied.
Na\"ively, $N_0$ is interpreted as the number of $e$-folds during the renormalization flow time scale. 
Furthermore, we denote the renormalization parameter  by $t$, which runs between $0$ (corresponding to $\tau_{\rm UV}$) and $t_{\rm tot}$ (corresponding to $\tau_{\rm IR}$): it represents the scale $\tau=\tau_{\rm UV}-t$.
The  value of $\zeta$ at $\tau$ subtracted from $\zeta_L$   is denoted by $\zeta_t$ (that is, $\zeta_t=\zeta_L-\zeta(\tau=\tau_{\rm UV}-t)$)).
Then by definition, $\zeta_{t=0}=0$ and $\zeta_{t=t_{\rm tot}}=\zeta_{t, {\rm tot}}$, respectively.

 Now, we consider the regularized  reheating surface volume  $e^{dh\zeta}$ in the form of
 \dis{O_{h, {\rm phys}}=e^{d h\zeta_L}f(L),}
 where $f(L)$ is required to make $O_{h, {\rm phys}}$ to be independent of the choice of $\epsilon_{\rm phys}$, hence $L$.
 Since
 \dis{\zeta_L=L+\tau_{\rm UV} = L+ \tau_{\rm IR}+t_{\rm tot},}
 where $L+\tau_{\rm IR}$ is fixed, the fluctuation of $\zeta_L$ is reflected in the fluctuation of $t_{\rm tot}$ only.
 Thus,  the coarse-graining (which will be simply denoted by $\langle \cdot \rangle$) of $\zeta_t$ for $ O_{h, {\rm phys}}$ is given by 
  \dis{O_{h, {\rm fid}}\equiv \langle O_{h, {\rm phys}}\rangle = f(L)e^{dh(L+\tau_{\rm IR})}\langle e^{d h t_{\rm tot}} \rangle,} 
and in the presence of  scale invariance (that is, $\varepsilon=0$), it becomes \cite{Cotler:2026lna} 
 \dis{O_{h, {\rm fid}} = :e^{d \alpha_h Q\zeta_{\rm IR}} :,\label{eq:Ohfid}}
 while $O_{h, {\rm phys}}$ can be written in the form of
 \dis{O_{h, {\rm phys}} = :e^{d \alpha_h Q\zeta_{\rm UV}} :.}

 To see this, we analyze the stochastic evolution of $\zeta_t$    by considering
 \dis{N(t)=N_0-t+\zeta_t.}
 That is, as $t$ increases from $t=0$ (where $\zeta_t=0$) to $t=t_{\rm tot}$ (where $\zeta_t=\zeta_{t, {\rm tot}}$), $N(t)$ undergoes a random walk along   $\zeta_t$ from $N_0$ to $0$ and then stops evolving. 
Given the value of $N_0$, $t_{\rm tot}$ is not a fixed value, but a quantity determined by the evolution of $N(t)$.
In the presence of scale invariance, which was analyzed in \cite{Cotler:2026lna}, $t_{\rm tot}$ follows the  probability distribution  
\dis{p(t_{\rm tot})=N_0 \sqrt{\frac{d Q^2}{4\pi t_{\rm tot}^3}}e^{-\frac{d Q^2(t_{\rm tot}-N_0)^2}{4t_{\rm tot}}}.\label{eq:prob}}
Compared with the Gaussian distribution, this probability distribution contains an additional $1/t_{\rm tot}$ factor.
While it is obtained by solving the Fokker-Planck equation \cite{ Creminelli:2008es, Dubovsky:2008rf}, one can understand this structure by noting that the running can be regarded as a first-passage process, where $N(t)$ never reaches $N=0$ for $t<t_{\rm tot}$ (for a review, see, e.g.,  \cite{Redner:2001}).
This is briefly sketched in Appendix \ref{App:ProbDist}.
Then we obtain
\dis{R_h (N_0) \equiv \langle e^{dht_{\rm tot}}\rangle=\int_0^\infty dt_{\rm tot} p(t_{\rm tot})e^{dht_{\rm tot}} =e^{d Q \alpha_h N_0},\label{eq:Rh}}
where 
\dis{\alpha_h=\frac{Q}{2}-\sqrt{\frac{Q^2}{4}-h},\label{eq:ah}}
i.e., $h=\alpha_h(Q-\alpha_h)$ is satisfied (see \eqref{eq:Delta1} with $\alpha=dQ\alpha_h$ and $\varepsilon=0$), which coincides with the value of $\alpha_h$ in \eqref{eq:ah1} for $\varepsilon=0$. 
Choosing $f(L)=e^{d a_h^2L}$ and using the relations $L+\tau_{\rm IR}+N_0=\zeta_{\rm IR}$, $h=\alpha_h(Q-\alpha_h)$,  as well as $\tau_{\rm IR}=[d Q^2/2]\langle \zeta_{\rm IR}^2\rangle$,
\footnote{This is obtained by replacing $r$ in \eqref{eq:G2} with $\varepsilon=0$ by $\hat{\epsilon}_{\rm IR}=e^{-\tau_{\rm IR}}$ and use the fact $\tau_{\rm IR} \gg k_{\rm min}$. }
 we arrive at \eqref{eq:Ohfid}:
\dis{  O_{h, {\rm fid}}&=e^{-d \alpha_h^2\tau_{\rm IR} } e^{d \alpha_h Q\zeta_{\rm IR}} =e^{-\frac{(d \alpha_h Q)^2}{2}\langle \zeta_{\rm IR}^2\rangle}e^{d \alpha_h Q\zeta_{\rm IR}}
\\
&=:e^{d \alpha_h Q\zeta_{\rm IR}}:.}

\subsubsection{Scaling dimension with the broken scale invariance}

 In our case, since  scale invariance is broken by nonzero $\varepsilon$, $\langle\zeta_{t, {\rm tot}}^2\rangle$ is not simply correlated with $t_{\rm tot}$.
 This makes it nontrivial to solve the Fokker-Planck equation to obtain the probability distribution, from which ensemble average $R_h(N_0, \tau_{\rm IR})$ in the form of \eqref{eq:Rh} is evaluated.
 \footnote{As can be inferred from \eqref{eq:varzeta}, the two point function with broken scale invariance  depends on the scale, which is reflected in an additional $\tau_{\rm IR}$ dependence of $R_h$.  }
 Then as an alternative approach, we use the fact that $\tau_{\rm IR}$ is given by the fixed value, while $\tau_{\rm UV}$ fluctuates along $\zeta_L$ (thus we need to take the ensemble average over it).
 More concretely, we obtain $\alpha_h$ by dividing the random walk from $N(t=0)=N_0$ to $N(t=t_{\rm tot})=0$ during $t_{\rm tot}$ into the last step during $\delta t$ to reach the fixed $\tau_{\rm IR}$ and the rest, i.e., the random walk from $\tau_{\rm UV}$ to $\tau_{\rm IR}+\delta t$ (where $t$ is given by $t'_{\rm tot} \equiv \tau_{\rm UV}-(\tau_{\rm IR}+\delta t)=t_{\rm tot}-\delta t$).
 We note that for the rest random walk, $\zeta'_{t, {\rm tot}}$, a value of $\zeta_t$ at $t'_{\rm tot}$, is given by $\zeta_L-\zeta(\tau_{\rm IR}+\delta t)$ by definition.
 Denoting $\zeta(\tau_{\rm IR}+\delta t)-\zeta_{\rm IR}=\zeta_{t, {\rm tot}}-\zeta'_{t, {\rm tot}}$ by $\delta \zeta$,
 \footnote{That is, since $\zeta_t=\zeta_L-\zeta(\tau=\tau_{\rm UV}-t)$, 
 \dis{\delta\zeta=\zeta(\tau_{\rm IR}+\delta t)-\zeta_{\rm IR}=-\zeta_L+\zeta(\tau_{\rm IR}+\delta t)+\zeta_L-\zeta_{\rm IR}=-\zeta'_{t, {\rm tot}}+\zeta_{t, {\rm tot}}. \nonumber} }
 one obtains $N(t_{\rm tot}-\delta t)=N_0-(t_{\rm tot}-\delta t)+\zeta'_{t, {\rm tot}}=\delta t-\delta \zeta$.
Since $\tau_{\rm IR}$ is fixed, $\tau_{\rm IR}+\delta t$ is fixed as well, then  the rest random walk is equivalent to the random walk on $N(t)$ over $N_0-\delta t+\delta \zeta$.
Meanwhile, the variance of $\delta\zeta$ during $\delta t$, which will be denoted by $\langle\delta\zeta^2\rangle$, is given by $\langle \zeta_{t, {\rm tot}}^2 \rangle - \langle {\zeta'_{t, {\rm tot}}}^2\rangle = \langle \zeta^2 (\tau_{\rm IR}+\delta t) \rangle - \langle \zeta_{\rm IR}^2\rangle$.
 After choosing $k_0=k_{\rm min}$, \eqref{eq:G2} leads to 
 \dis{\langle \zeta^2 (\tau_{\rm IR}+\delta t) \rangle &=\frac{2}{d Q^2(k_{\rm min})}\log\Big(\frac{1}{k_{\rm min}  e^{-(\tau_{\rm IR}+\delta t)}  }\Big)\Big[1- \frac{\varepsilon}{2}\log(k_{\rm min} e^{-(\tau_{\rm IR}+\delta t)}) +{\cal O}(\varepsilon^2)\Big]
\\
&=\frac{2}{d Q^2(k_{\rm min})}\Big[(\tau_{\rm IR}+\delta t)+\frac{\varepsilon}{2}(\tau_{\rm IR}+\delta t)^2 +{\rm constant}+{\cal O}(\varepsilon^2)\Big]
\\
&=\langle\zeta_{\rm IR}^2\rangle +\frac{2}{d Q^2 }\big(1+\varepsilon \tau_{\rm IR}\big)\delta t +{\cal O}(\varepsilon^2, \delta t^2), \label{eq:z2t}}
where we briefly denote $Q(k_{\rm min})$ by $Q$.
Therefore, 
\dis{\langle\delta\zeta^2\rangle=\frac{2}{d Q^2 }\big(1+\varepsilon \tau_{\rm IR}\big)\delta t,\label{eq:z2tau}}
which is the same form as \eqref{eq:varzeta} with $\zeta_1$ replaced by $\tau_{\rm IR}$, up to the choice of the renormalization scale $k_0$.  
 Indeed, while $\zeta_1$ and $\zeta_2$ used to obtain \eqref{eq:varzeta} satisfy $\epsilon=e^{\zeta_1}\hat{\epsilon}_1=e^{\zeta_2} \hat{\epsilon_2}$ with $\epsilon$ fixed, one may also use the parametrization $\hat{\epsilon}_{1,2}=e^{-\tau_{1,2}}$, similarly to the approach used for \eqref{eq:z2tau}. 
 In this case, since $\epsilon$ is fixed, $\delta t\equiv \tau_1-\tau_2$ is identical to $\delta \zeta=\zeta_1-\zeta_2$, hence,  $\zeta_1$ and $\delta \zeta$ in \eqref{eq:varzeta} may be replaced by $\tau_1$ and $\delta t$, respectively.
 Then \eqref{eq:varzeta} can be written in the exactly the same form as \eqref{eq:z2tau}.

  Now, taking into account  the separation of  the last step from the rest and denoting the derivative with respect to $t$ and $N_0$ by a dot and a prime, respectively, $R_h(N_0)\equiv \langle e^{dh t_{\rm tot}}\rangle$ can be rewritten as 
 {\small
  \dis{R_h(N_0, \tau_{\rm IR})\equiv \langle e^{dh t_{\rm tot}}\rangle &= \langle e^{dh(\delta t+t'_{\rm tot})}\rangle = e^{dh\delta t}R_h(N'_0, \tau_{\rm IR}+\delta t)
  \\
  =(1+ dh\delta t)\Big[&R_h(N_0, \tau_{\rm IR})+\langle(N'_0-N_0)\rangle R'_h(N_0, \tau_{\rm IR})+\delta t \dot{R}_h(N_0, \tau_{\rm IR})
\\
&+\frac12\langle (N'_0-N_0)^2\rangle R''_h(N_0, \tau_{\rm IR})+\cdots\Big]
  \\
  =(1+ dh\delta t)\Big[& R_h(N_0, \tau_{\rm IR})+\langle(\delta\zeta-\delta t)\rangle R'_h(N_0, , \tau_{\rm IR})+\delta t \dot{R}_h(N_0, \tau_{\rm IR})
  \\
  &+\frac12\langle (\delta\zeta-\delta t)^2\rangle R''_h(N_0, , \tau_{\rm IR})+\cdots\Big].} 
 Then from 
\dis{\langle\delta \zeta \rangle=0,\quad\quad \langle\delta \zeta^2\rangle=\frac{2}{dQ^2}(1+\varepsilon\tau_{\rm IR})\delta t,\label{eq:zetainif}} 
}
  it becomes 
  {\small
 \dis{R_h(N_0, \tau_{\rm IR})&\simeq (1+ dh\delta t) \Big[R_h(N_0, \tau_{\rm IR})-\delta t R'_h(N_0, \tau_{\rm IR})+\delta t \dot{R}_h(N_0, \tau_{\rm IR})+\frac{1+\varepsilon\tau_{\rm IR}}{dQ^2} \delta t R''_h(N_0, \tau_{\rm IR}) \Big]
 \\
 &=R_h(N_0, \tau_{\rm IR})+\delta t\Big[dh R_h(N_0, \tau_{\rm IR}) -R'_h(N_0, \tau_{\rm IR})+\dot{R}_h(N_0, \tau_{\rm IR})+ \frac{1+\varepsilon\tau_{\rm IR}}{dQ^2}R''_h(N_0, \tau_{\rm IR})\Big]
 \\
 &\quad+{\cal O}(\delta t^2),}
 }
which leads to
 \dis{\dot{R}_h(N_0, \tau_{\rm IR})+dh R_h(N_0, \tau_{\rm IR}) -R'_h(N_0, \tau_{\rm IR})+ \frac{1+\varepsilon\tau_{\rm IR}}{dQ^2}R''_h(N_0, \tau_{\rm IR})=0.\label{eq:Eq}}
As a boundary condition, we require that $R_h=1$ for $N_0=0$ (which means that $\tau_{\rm UV}=\tau_{\rm IR}$).

 We solve equation \eqref{eq:Eq} using a perturbative expansion in $\varepsilon$.
 For this purpose, we employ an ansatz
 \dis{R_h(N_0, \tau_{\rm IR})=e^{\alpha(\tau_{\rm IR})N_0+\varepsilon\beta(N_0, \tau_{\rm IR})+{\cal O}(\varepsilon^2)}.}
 We note that since the $\tau_{\rm IR}$ dependence arises from   broken scale invariance, which is evident from the fact that  $\tau_{\rm IR}$ in \eqref{eq:Eq} appears multiplied by $\varepsilon$, the derivative with respect to $\tau_{\rm IR}$ raises the order in $\varepsilon$ by one: $\dot{\alpha}$ and $\varepsilon \dot{\beta}$ are of order of ${\cal O}(\varepsilon)$ and ${\cal O}(\varepsilon^2)$, respectively.
For this reason, we can apply an adiabatic approximation, such that $\alpha(\tau_{\rm IR})$ is given by $dQ \alpha_h$, where
 \dis{\alpha_h(\tau_{\rm IR})=\frac{1}{2\sqrt{1+\varepsilon\tau_{\rm IR}}}\Bigg[\frac{Q}{\sqrt{1+\varepsilon\tau_{\rm IR}}}-\sqrt{\frac{Q^2}{1+\varepsilon\tau_{\rm IR}}-4h} ~\Bigg],  \label{eq:ahtIR}}
 such that 
 \dis{d h -\alpha(\tau_{\rm IR})+\frac{1+\varepsilon\tau_{\rm IR}}{dQ^2}\alpha^2(\tau_{\rm IR})=0\label{eq:ex1}}
 is satisfied.
 Comparing with \eqref{eq:ah}, one finds that $Q$ is replaced by
 \footnote{Indeed,  $\langle \delta\zeta^2\rangle$ in \eqref{eq:zetainif} also can be simply written as $[2/(d Q_\tau^2)]\delta t$.}
 \dis{Q_\tau \equiv\frac{Q}{\sqrt{1+\varepsilon\tau}}.}
We also note that differentiating \eqref{eq:ex1} with respect to $\tau_{\rm IR}$ gives
\dis{ \dot{\alpha} (\tau_{\rm IR}) =\frac{\varepsilon}{dQ^2} \frac{1}{\sqrt{1-\frac{4 h}{Q^2}(1+\varepsilon \tau_{\rm IR})}}{\alpha}^2 (\tau_{\rm IR}). \label{eq:ex2} }
Then \eqref{eq:Eq} can be written as an ${\cal O}(\varepsilon)$ equation for $\beta$:
 \dis{\dot{\alpha}(\tau_{\rm IR}) N_0-\varepsilon\sqrt{1-\frac{4 h}{Q^2}(1+\varepsilon \tau_{\rm IR})}~\beta'(\tau_{\rm IR})+\varepsilon\frac{1+\varepsilon\tau_{\rm IR}}{d Q^2}\beta''(\tau_{\rm IR})=0.}
 Expanding $\beta'(\tau_{\rm IR})$ as $c_0(\tau_{\rm IR})+c_1(\tau_{\rm IR}) N_0$ (thus $\beta''(\tau_{\rm IR})=c_1(\tau_{\rm IR})$ and $\beta(\tau_{\rm IR})=c_0(\tau_{\rm IR})N_0+\frac12 c_1(\tau_{\rm IR}) N_0^2$ in accordance with the boundary condition $R_h(N_0=0, \tau_{\rm IR})=1$), these equations give
 \dis{{\cal O}((N_0)^0)\quad &: \quad -\sqrt{1-\frac{4 h}{Q^2}(1+\varepsilon \tau_{\rm IR})}~ c_0(\tau_{\rm IR})+\frac{1+\varepsilon\tau_{\rm IR}}{d Q^2}c_1(\tau_{\rm IR})=0,
 \\
{\cal O}((N_0)^1)\quad &: \quad\dot{\alpha}(\tau_{\rm IR})-\varepsilon\sqrt{1-\frac{4 h}{Q^2}(1+\varepsilon \tau_{\rm IR})}~ c_1(\tau_{\rm IR}) =0, }
at each order in $N_0$.
Referring to \eqref{eq:ex2},  one obtains
\dis{c_0(\tau_{\rm IR}) &=\frac{ (1+\varepsilon\tau_{\rm IR})}{(dQ^2)^2} \frac{1}{ \Big( 1-\frac{4 h}{Q^2}(1+\varepsilon \tau_{\rm IR})\Big)^{3/2}}{\alpha}^2 (\tau_{\rm IR}),
\\
c_1(\tau_{\rm IR}) &=\frac{1}{dQ^2} \frac{1}{ 1-\frac{4 h}{Q^2}(1+\varepsilon \tau_{\rm IR})}{\alpha}^2 (\tau_{\rm IR}).}

In summary,  $\log R_h(N_0, \tau_{\rm IR})$ corrected up to ${\cal O}(\varepsilon)$ is given by
\dis{\log R_h(N_0, \tau_{\rm IR}) =& \alpha(\tau_{\rm IR})N_0
+ \frac{\varepsilon (1+\varepsilon\tau_{\rm IR})}{(dQ^2)^2} \frac{1}{ \Big( 1-\frac{4 h}{Q^2}(1+\varepsilon \tau_{\rm IR})\Big)^{3/2}}{\alpha}^2 (\tau_{\rm IR}) N_0
\\
&+\frac12\frac{\varepsilon}{dQ^2} \frac{1}{ 1-\frac{4 h}{Q^2}(1+\varepsilon \tau_{\rm IR})}{\alpha}^2 (\tau_{\rm IR}) N_0^2 +{\cal O}(\varepsilon^2),}
with 
\dis{\alpha(\tau_{\rm IR}) = &\frac{d Q}{2\sqrt{1+\varepsilon\tau_{\rm IR}}}\Bigg[\frac{Q}{\sqrt{1+\varepsilon\tau_{\rm IR}}}-\sqrt{\frac{Q^2}{1+\varepsilon\tau_{\rm IR}}-4h} ~\Bigg]
\\
\equiv &\frac{d Q_\tau}{2 }\Big[Q_\tau-\sqrt{Q_\tau^2-4h} ~\Big].\label{eq:ah2}}
Furthermore, the adiabatic behavior of $R_h(N_0, \tau_{\rm IR})$ with respect to $\tau_{\rm IR}$, which can be inferred from \eqref{eq:ex1}, implies that 
\dis{\Delta=dh=\alpha-\frac{\alpha^2}{dQ^2}(1+\varepsilon \tau_{\rm IR})+{\cal O}(\varepsilon)=\alpha-\frac{\alpha^2}{d Q_\tau^2}+{\cal O}(\varepsilon),\label{eq:Delta}}
is a good approximation for the scaling dimension.
This is consistent with  \eqref{eq:Delta1}.

 \section{Discussion}
 \label{Sec:Discussion}
 
  In this section, we discuss   physical meaning of the corrected scaling dimension given by  \eqref{eq:ah2} and \eqref{eq:Delta}, focusing on the replacement of $Q$ by $Q_\tau$.
  First, the condition for $\alpha$ (hence $\alpha_h$) to take a real value,  interpreted as the renormalizability   of the reheating surface volume, is given by
 \dis{Q_\tau^2=\frac{Q^2}{1+\varepsilon\tau_{\rm IR}} \geq 4h,\quad\quad {\rm or}\quad\quad\epsilon(k_0) \geq \frac{3}{4\pi^2}\frac{H^2(k_0)}{M_{\rm Pl}^2}(1+\varepsilon \tau_{\rm IR}),\label{eq:Cond}}  
where for the last inequality we set $d=3$ and $h=1$. 
Therefore, when $\varepsilon=n_s-1<0$, $Q_\tau$ gets larger as $\tau_{\rm IR}$ increases (i.e., as the scale becomes more macroscopic), making the reality condition for $\alpha$ easier to be   satisfied.  
At the same time, the fluctuation at the last step   $\langle\delta \zeta^2\rangle$ ($\propto 1/Q_\tau^2$)  gets smaller.
This shows  how  breaking of scale invariance by the negative $\varepsilon$ leads to macroscopically smooth reheating surface, which is reflected in $\alpha \in \mathbb{R}$, or equivalently, the well-defined renormalized reheating surface volume $e^{\alpha \zeta}$ (=$e^{dQ\alpha_h\zeta}$).
Even though  the reheating surface may wrinkle at the microscopic scale, this wrinkling is not maintained at the macroscopic scale,  as  the fractal pattern is lost along with broken scale invariance.

 Indeed, as is evident from the definition \eqref{eq:ns-1},  the condition $\varepsilon<0$ means that the amplitude ${\cal P}_\zeta(k)$ decreases as $k$ increases. 
Since the  higher $k$-mode  crosses  the horizon later, it corresponds to the larger value of the time fluctuation $\zeta$ (recalling from  \eqref{eq:metric} that $\zeta$ is interpreted as the fluctuation of $e$-folding number in dS limit), thus more macroscopic scale.
Therefore, $\varepsilon<0$ is interpreted as the decrease of ${\cal P}_\zeta(k)$ on macroscopic scale, which is consistent with the behavior of the reheating surface described above.
This is also evident from the specific case  realizing $\varepsilon= n_s-1=-2\epsilon-\tilde{\eta}<0$: even though $\varepsilon$  is very tiny, if $\tilde{\eta}=\dot{\epsilon}/(H\epsilon)>0$, $\epsilon$ (which is positive) increases in time, then $|\varepsilon|$, thus $Q_\tau^2$ get larger as time goes on.
Therefore, eventually, at the reheating moment, \eqref{eq:Cond} is easy to be satisfied.
We also note that the negative value of $\varepsilon$ is consistent with observations: recent Planck 2018 results provide  \cite{Planck:2018jri}
\dis{n_s=0.9649 \pm 0.0042,\quad\quad {\rm thus}\quad\quad \varepsilon=n_s-1=-0.0351 \pm 0.0042.}

On the other hand, as we have seen in Section \ref{subsec:appr2}, when $|\varepsilon|\ll 1$, $Q_\tau$ varies slowly with respect to $\tau_{\rm  IR}$, allowing the adiabatic approximation.
Then the probability distribution may be approximated by that given by \eqref{eq:prob} (i.e., that for $\varepsilon=0$) with $Q$ replaced by $Q_\tau$.
Then for $t_{\rm tot} \gg 1$, the integration \eqref{eq:Rh} behaves as (see also discussion in \cite{Creminelli:2008es, Cotler:2026lna})
\dis{\int d t_{\rm tot} p(t_{\rm tot})e^{dht_{\rm tot}} \sim \int dt_{\rm tot} \frac{1}{t_{\rm tot}^{3/2}}e^{-d\big(\frac{Q_\tau^2}{4}-h\big)t_{\rm tot}},}
which converges provided the renormalizability condition $Q_\tau^2\geq 4$ is satisfied.
Indeed, this shows explicitly that the increase of $Q_\tau$ suppresses the probability for the large fluctuation $\zeta_{t, {\rm tot}} \gg 1$, such that the volume expansion effect $e^{d h t_{\rm tot}}$ is compensated.
Since the total exponent is negative, the integration converges.

We also note that when $\tau_{\rm IR}$ increases to approach $-1/\varepsilon$ from below,
 \footnote{We note that from the expression \eqref{eq:z2tau} for $\langle\delta \zeta^2\rangle$ which is positive,    $1+\varepsilon\tau_{\rm IR} $ is required to be positive as well.}
 $|\varepsilon\tau_{\rm IR}|$ cannot be treated as a small value any longer, hence higher order corrections of ${\cal O}((\varepsilon \tau_{\rm IR})^2)$ must be taken into account as well.
 While these higher order terms   are not explicitly considered   in our discussion, we expect that the breaking effect of scaling invariance becomes more evident in the $\tau_{\rm IR} \to -1/\varepsilon$ limit, i.e., the macroscopic limit, thereby the reheating surface volume can be described classically.
 This  expected behavior is  already captured in the  ${\cal O}(\varepsilon\tau_{\rm IR})$ terms as well as the expression of $Q_\tau$, which can be regarded as a resummation of corrections to $Q$.
 First,  as $\langle\delta \zeta^2\rangle$ given by \eqref{eq:z2tau} vanishes, the fluctuation effect  also becomes negligible, from which we expect that the reheating surface volume behaves classically.
 Moreover, not only the value of $Q_\tau$, but also the increasing rate $d\log Q_\tau^2/d\tau_{\rm IR}=-\varepsilon/(1+\varepsilon\tau_{\rm IR})$ (which is positive since $\varepsilon<0$) diverges, indicating the condition $Q_\tau^2 \geq 4h$ is trivially satisfied. 
  This also can be found from the value of $\alpha_h$ in the $\tau_{\rm IR}\to -1/\varepsilon$ limit: for $\tau_{\rm IR}\simeq  -1/\varepsilon$, the second term $-[\alpha^2/(dQ^2)](1+\varepsilon \tau_{\rm IR})$ in  \eqref{eq:Delta} which originates from $\langle\delta \zeta^2\rangle$ vanishes, just satisfying $h=\alpha=Q\alpha_h$.
In addition,  $\alpha_h$ in \eqref{eq:ahtIR} is approximated by $h/Q$, then $O_{h, {\rm fid}}=:e^{d\alpha_h Q\zeta_{\rm IR}}:$ is just given by $:e^{d h\zeta_{\rm IR}}:$.
 Regarding this, we note that while \eqref{eq:Delta} allows another solution
 \dis{\alpha_{h,+}=\frac{1}{2(1+\varepsilon\tau_{\rm IR})}\big[Q+\sqrt{Q^2-4h(1+\varepsilon\tau_{\rm IR})}\big],}
 it is difficult to  be regarded as a physically meaningful value for the reheating surface volume operator.
 It is because for  $\tau_{\rm IR}= -1/\varepsilon-\delta\tau$, one obtains $\alpha_{h,+}\simeq -Q/(\varepsilon\delta\tau)$, which diverges in the $\delta\tau \to \infty$ limit, thus the operator $:e^{d\alpha_{h,+} Q\zeta_{\rm IR}}:$ diverges.
 In fact, when $\varepsilon=0$, \eqref{eq:Delta} can be written as $h=\alpha_h(Q-\alpha_h)$, and it is invariant under the exchange $\alpha_h \leftrightarrow Q-\alpha_h$, which leads to $\alpha_h \leftrightarrow \alpha_{h, +}$ where $\alpha_h$ is given by \eqref{eq:ah}.
 In this case, the operator $:e^{d\alpha_{h,+} Q\zeta_{\rm IR}}:$ can be treated as equivalent to $:e^{d\alpha_h Q\zeta_{\rm IR}}:$.
In contrast, for $\varepsilon \ne 0$, the exchange symmetry is broken, then one can regard   $\alpha_{h,+}$ as the scaling dimension of an independent operator.
However, since it diverges as $\tau_{\rm IR}$ increases, it may not be appropriate to be interpreted as a well-behaving scaling dimension at the macroscopic scale, even though the divergence may be reduced by ${\cal O}((\varepsilon\tau_{\rm IR})^2)$ effects.

 \section{Conclusions}
\label{sec:conclusion}

 In this article, we explored how the effects of broken scale invariance modify the no-eternal inflation condition following an approach suggested in \cite{Cotler:2026lna}, where the condition is interpreted as the renormalizability condition for the reheating surface volume in the presence of scale invariance.
 More concretely, the scaling dimension for the reheating surface volume operator is well-defined provided the no-eternal inflation condition is satisfied.
 By evaluating the scaling dimension by taking into account the broken scale invariance parametrized by the spectral index, we show that so far as $\varepsilon = n_s-1$ is negative, it moderates the condition for the failure of renormalization.
 This is interpreted that the breaking of scale invariance by the negative $\varepsilon$ reduces the fluctuation reflected in the  wrinkling of the reheating surface volume as the scale becomes  macroscopic.

 While our conclusion was also expected in previous works, such as \cite{Dubovsky:2008rf, Lewandowski:2013aka}, our approach differs from them in two aspects.
 First, instead of solving the Fokker-Planck equation numerically, we use the renormalization approach of \cite{Cotler:2026lna}, to provide a more transparent   physical interpretation.
 Second, we did not assume any specific potential, by using model independent parametrization of the scale invariance breaking.
 However, our analysis remains at the leading order approximation, which motivates   future work to study higher order effects more systematically with appropriate interpretation and compare them with the previous studies.

%\subsection*{Acknowledgements}

%

%\newpage

\appendix

\section{Understanding probability distribution}
\label{App:ProbDist}

In this Appendix, we discuss how the probability distribution given by \eqref{eq:prob} can be obtained.
Since $N(t)$ is not required to reach zero for $t<t_{\rm tot}$, the evolution, or equivalently, the renormalization running, of $N(t)$ can be interpreted as the first passage process.
To see the situation intuitively, consider a molecule   undergoing a random walk from $x=N_0$ to $x=0$ where it is absorbed.
In this $1$-dimensional motion, the molecule  moves in discrete steps of size $\sigma$ to the left (toward $x=0$) and right (toward $x=N_0$) randomly before being absorbed.
Then the interval $[0, N_0]$ consists of $k \equiv N_0/\sigma$ steps.
 Denoting the total number of steps taken by the molecule by $n$, it is given by the sum of $n_+$, the number of steps in the right direction, and  $n_-$, the number of steps in the left direction.
 That is,
 \dis{&n_++n_-=n,\quad n_+-n_-=-\frac{N_0}{\sigma}=-k,\quad\quad {\rm thus}
 \\
 &n_+=\frac{n-k}{2},\quad n_-=\frac{n+k}{2}}
 are satisfied.

 Now, for the first passage process, the molecule must move to the left at the last step just before reaching $x=0$, the number of possible ways for which is given by
 \dis{\binom{n-1}{\frac{(n-1)+(k-1)}{2}} = \binom{n-1}{\frac{n+k}{2}-1},}
since the molecule travels the interval $[0, N_0-\sigma=(k-1)\sigma]$ with $(n-1)$ steps before the last step.
But still, it also includes the path where the molecule reaches $x=0$ before $n-1$ steps, which can be identified with the random walk from $x=-N_0$ to $x=\sigma$ (the number of steps for this interval is given by $k+1$, since $N_0+\sigma=(k+1)\sigma$) in $n-1$ steps.
Since the number of possible ways for it is given by
\dis{\binom{n-1}{\frac{(n-1)+(k+1)}{2}} = \binom{n-1}{\frac{n+k}{2}},}
The number of the first passage processes is given by
\dis{ \binom{n-1}{\frac{n+k}{2}-1}-\binom{n-1}{\frac{n+k}{2}} =\frac{k}{n} \binom{n}{\frac{n+k}{2}}.}
Therefore, denoting the probability for the molecule to move to the left by $p$, the probability distribution is given by
\dis{\frac{k}{n} \binom{n}{\frac{n+k}{2}} p^{\frac{n+k}{2}}(1-p)^{\frac{n-k}{2}}.\label{eq:pr(1)}} 
 In the continuum limit of this distribution where $n$ is very large, the binomial part is approximated by the Gaussian distribution, where the mean value and the variance of the location of the molecule after $n$ steps are given by 
 \dis{\langle x_n\rangle&=N_0 + n \big(p\times(-\sigma)+(1-p)\times\sigma\big)=N_0-n\sigma(2p-1),
 \\
 \sigma_x^2&=n \Big[\big(p\times(-\sigma)^2+(1-p)\times\sigma^2\big)-\big(\sigma(2p-1)\big)^2\Big]
=n \sigma^2\big(1-(2p-1)^2\big),}
respectively.

 As for $p$, one can set
 \dis{p=\frac12\Big(1+\frac{v\delta t}{\sigma}\Big),}
 such that $\langle x_n\rangle=N_0-n v\delta t = N_0-t_{\rm tot}$ with $v=1$.
While $v\delta t/\sigma =\delta t/\sigma \leq 1$ to make $p\leq 1$, we can make $\delta t$ to be infinitesimally small such that $v\delta t=\delta t \ll \sigma$, thus $\sigma_x^2\simeq n\sigma^2$ is satisfied.
In the presence of  scale invariance, $\sigma^2$ is identified with    \eqref{eq:z2tau} with $\varepsilon=0$, i.e., $\langle\delta\zeta^2\rangle=[2/dQ^2]\delta t$ for every step, from which one obtains   $\sigma_x^2=[2/dQ^2]n\delta t=[2/dQ^2]t_{\rm tot}$.
Then \eqref{eq:pr(1)} is written as
\dis{\frac{k}{n}  \sqrt{\frac{d Q^2}{4\pi t_{\rm tot}}}e^{-\frac{d Q^2(t_{\rm tot}-N_0)^2}{4t_{\rm tot}}} \sigma,}
where $\sigma$ is identified with $dx$.
On the other hand, since $k=N_0/\sigma$ and $n= t_{\rm tot}/\delta t$, the ratio $k/n$ is given by $(N_0/t)(\delta t/\sigma)$, where $\delta t/\sigma$ replaces $dx=\sigma$ by $dt_{\rm tot}=\delta t$.
Therefore, we arrive at
\dis{ N_0  \sqrt{\frac{d Q^2}{4\pi t_{\rm tot}^3}}e^{-\frac{d Q^2(t_{\rm tot}-N_0)^2}{4t_{\rm tot}}} d t_{\rm tot},}
which coincides with the probability density given by \eqref{eq:prob}.

\renewcommand{\theequation}{\Alph{section}.\arabic{equation}}

%\section{Uncertainty for the infrared modes}
%\label{app:IRuncert}
%\setcounter{equation}{0}


\begin{thebibliography}{99}

\small

%\cite{Steinhardt:1982kg}
\bibitem{Steinhardt:1982kg}
P.~J.~Steinhardt,
``Natural inflation,''
in The Very Early Universe, Proceedings of the
Nuffield Workshop, Cambridge, 21 June- 9 July, 1982, eds: G. W. Gibbons, S. W. Hawking
and S. T. C. Siklos (Cambridge University Press),
UPR-0198T.
%13 citations counted in INSPIRE as of 23 Sep 2026

%\cite{Vilenkin:1983xq}
\bibitem{Vilenkin:1983xq}
A.~Vilenkin,
%``The Birth of Inflationary Universes,''
Phys. Rev. D \textbf{27}, 2848 (1983)
%doi:10.1103/PhysRevD.27.2848
%1142 citations counted in INSPIRE as of 23 Sep 2026

%\cite{Linde:1986fc}
\bibitem{Linde:1986fc}
A.~D.~Linde,
%``ETERNAL CHAOTIC INFLATION,''
Mod. Phys. Lett. A \textbf{1}, 81 (1986)
%doi:10.1142/S0217732386000129
%367 citations counted in INSPIRE as of 23 Sep 2026

%\cite{Linde:1986fd}
\bibitem{Linde:1986fd}
A.~D.~Linde,
%``Eternally Existing Selfreproducing Chaotic Inflationary Universe,''
Phys. Lett. B \textbf{175}, 395-400 (1986)
%doi:10.1016/0370-2693(86)90611-8
%886 citations counted in INSPIRE as of 23 Sep 2026

%\cite{Goncharov:1987ir}
\bibitem{Goncharov:1987ir}
A.~S.~Goncharov, A.~D.~Linde and V.~F.~Mukhanov,
%``The Global Structure of the Inflationary Universe,''
Int. J. Mod. Phys. A \textbf{2}, 561-591 (1987)
%doi:10.1142/S0217751X87000211
%378 citations counted in INSPIRE as of 23 Sep 2026

%\cite{Guth:2007ng}
\bibitem{Guth:2007ng}
A.~H.~Guth,
%``Eternal inflation and its implications,''
J. Phys. A \textbf{40}, 6811-6826 (2007)
%doi:10.1088/1751-8113/40/25/S25
[arXiv:hep-th/0702178 [hep-th]].
%411 citations counted in INSPIRE as of 23 Sep 2026

%\cite{Starobinsky:1986fx}
\bibitem{Starobinsky:1986fx}
A.~A.~Starobinsky,
``Stochastic De Sitter (Inflationary) Stage in the Early Universe,''
Lect. Notes Phys. \textbf{246}, 107-126 (1986)
%doi:10.1007/3-540-16452-9{\_}6
%554 citations counted in INSPIRE as of 23 Sep 2026

%\cite{Starobinsky:1994bd}
\bibitem{Starobinsky:1994bd}
A.~A.~Starobinsky and J.~Yokoyama,
%``Equilibrium state of a selfinteracting scalar field in the De Sitter background,''
Phys. Rev. D \textbf{50}, 6357-6368 (1994)
%doi:10.1103/PhysRevD.50.6357
[arXiv:astro-ph/9407016 [astro-ph]].
%911 citations counted in INSPIRE as of 23 Sep 2026



%\cite{Cruces:2022imf}
\bibitem{Cruces:2022imf}
D.~Cruces,
%``Review on Stochastic Approach to Inflation,''
Universe \textbf{8}, no.6, 334 (2022)
%doi:10.3390/universe8060334
[arXiv:2203.13852 [gr-qc]].
%56 citations counted in INSPIRE as of 23 Sep 2026

%\cite{Aryal:1987vn}
\bibitem{Aryal:1987vn}
M.~Aryal and A.~Vilenkin,
%``The Fractal Dimension of Inflationary Universe,''
Phys. Lett. B \textbf{199}, 351-357 (1987)
%doi:10.1016/0370-2693(87)90932-4
%107 citations counted in INSPIRE as of 24 Sep 2026

%\cite{Winitzki:2001np}
\bibitem{Winitzki:2001np}
S.~Winitzki,
%``The Eternal fractal in the universe,''
Phys. Rev. D \textbf{65}, 083506 (2002)
%doi:10.1103/PhysRevD.65.083506
[arXiv:gr-qc/0111048 [gr-qc]].
%38 citations counted in INSPIRE as of 24 Sep 2026

%\cite{Winitzki:2008ph}
\bibitem{Winitzki:2008ph}
S.~Winitzki,
%``Reheating-volume measure for random-walk inflation,''
Phys. Rev. D \textbf{78}, 063517 (2008)
%doi:10.1103/PhysRevD.78.063517
[arXiv:0805.3940 [gr-qc]].
%16 citations counted in INSPIRE as of 23 Sep 2026

%\cite{Creminelli:2008es}
\bibitem{Creminelli:2008es}
P.~Creminelli, S.~Dubovsky, A.~Nicolis, L.~Senatore and M.~Zaldarriaga,
%``The Phase Transition to Slow-roll Eternal Inflation,''
JHEP \textbf{09}, 036 (2008)
%doi:10.1088/1126-6708/2008/09/036
[arXiv:0802.1067 [hep-th]].
%111 citations counted in INSPIRE as of 24 Sep 2026

%\cite{Dubovsky:2008rf}
\bibitem{Dubovsky:2008rf}
S.~Dubovsky, L.~Senatore and G.~Villadoro,
%``The Volume of the Universe after Inflation and de Sitter Entropy,''
JHEP \textbf{04}, 118 (2009)
%doi:10.1088/1126-6708/2009/04/118
[arXiv:0812.2246 [hep-th]].
%51 citations counted in INSPIRE as of 24 Sep 2026

%\cite{Dubovsky:2011uy}
\bibitem{Dubovsky:2011uy}
S.~Dubovsky, L.~Senatore and G.~Villadoro,
%``Universality of the Volume Bound in Slow-Roll Eternal Inflation,''
JHEP \textbf{05}, 035 (2012)
%doi:10.1007/JHEP05(2012)035
[arXiv:1111.1725 [hep-th]].
%41 citations counted in INSPIRE as of 24 Sep 2026

%\cite{Matsui:2018bsy}
\bibitem{Matsui:2018bsy}
H.~Matsui and F.~Takahashi,
%``Eternal Inflation and Swampland Conjectures,''
Phys. Rev. D \textbf{99}, no.2, 023533 (2019)
%doi:10.1103/PhysRevD.99.023533
[arXiv:1807.11938 [hep-th]].
%100 citations counted in INSPIRE as of 25 Sep 2026

%\cite{Dimopoulos:2018upl}
\bibitem{Dimopoulos:2018upl}
K.~Dimopoulos,
%``Steep Eternal Inflation and the Swampland,''
Phys. Rev. D \textbf{98}, no.12, 123516 (2018)
%doi:10.1103/PhysRevD.98.123516
[arXiv:1810.03438 [gr-qc]].
%72 citations counted in INSPIRE as of 25 Sep 2026

%\cite{Kinney:2018kew}
\bibitem{Kinney:2018kew}
W.~H.~Kinney,
%``Eternal Inflation and the Refined Swampland Conjecture,''
Phys. Rev. Lett. \textbf{122}, no.8, 081302 (2019)
%doi:10.1103/PhysRevLett.122.081302
[arXiv:1811.11698 [astro-ph.CO]].
%102 citations counted in INSPIRE as of 25 Sep 2026

%\cite{Brahma:2019iyy}
\bibitem{Brahma:2019iyy}
S.~Brahma and S.~Shandera,
%``Stochastic eternal inflation is in the swampland,''
JHEP \textbf{11}, 016 (2019)
%doi:10.1007/JHEP11(2019)016
[arXiv:1904.10979 [hep-th]].
%44 citations counted in INSPIRE as of 25 Sep 2026

%\cite{Rudelius:2019cfh}
\bibitem{Rudelius:2019cfh}
T.~Rudelius,
%``Conditions for (No) Eternal Inflation,''
JCAP \textbf{08}, 009 (2019)
%doi:10.1088/1475-7516/2019/08/009
[arXiv:1905.05198 [hep-th]].
%98 citations counted in INSPIRE as of 25 Sep 2026

%\cite{Wang:2019eym}
\bibitem{Wang:2019eym}
Z.~Wang, R.~Brandenberger and L.~Heisenberg,
%``Eternal Inflation, Entropy Bounds and the Swampland,''
Eur. Phys. J. C \textbf{80}, no.9, 864 (2020)
%doi:10.1140/epjc/s10052-020-8412-x
[arXiv:1907.08943 [hep-th]].
%42 citations counted in INSPIRE as of 25 Sep 2026

%\cite{Seo:2021bpb}
\bibitem{Seo:2021bpb}
M.~S.~Seo,
%``Implication of the swampland distance conjecture on the Cohen{\textendash}Kaplan{\textendash}Nelson bound in de Sitter space,''
Eur. Phys. J. C \textbf{82}, no.4, 338 (2022)
%doi:10.1140/epjc/s10052-022-10321-0
[arXiv:2106.00138 [hep-th]].
%11 citations counted in INSPIRE as of 25 Sep 2026%\cite{Seo:2021bpb}

%\cite{Seo:2022uaz}
\bibitem{Seo:2022uaz}
M.~S.~Seo,
%``Black hole production, eternal inflation, and information in quasi-de~Sitter space,''
JCAP \textbf{11}, 005 (2022)
%doi:10.1088/1475-7516/2022/11/005
[arXiv:2206.05857 [hep-th]].
%5 citations counted in INSPIRE as of 25 Sep 2026




%\cite{Arkani-Hamed:2007ryv}
\bibitem{Arkani-Hamed:2007ryv}
N.~Arkani-Hamed, S.~Dubovsky, A.~Nicolis, E.~Trincherini and G.~Villadoro,
%``A Measure of de Sitter entropy and eternal inflation,''
JHEP \textbf{05}, 055 (2007)
%doi:10.1088/1126-6708/2007/05/055
[arXiv:0704.1814 [hep-th]].
%210 citations counted in INSPIRE as of 25 Sep 2026

%\cite{Danielsson:2018ztv}
\bibitem{Danielsson:2018ztv}
U.~H.~Danielsson and T.~Van Riet,
%``What if string theory has no de Sitter vacua?,''
Int. J. Mod. Phys. D \textbf{27}, no.12, 1830007 (2018)

%doi:10.1142/S0218271818300070
[arXiv:1804.01120 [hep-th]].
%478 citations counted in INSPIRE as of 25 Sep 2026

%\cite{Obied:2018sgi}
\bibitem{Obied:2018sgi}
G.~Obied, H.~Ooguri, L.~Spodyneiko and C.~Vafa,
%``De Sitter Space and the Swampland,''
[arXiv:1806.08362 [hep-th]].
%1165 citations counted in INSPIRE as of 25 Sep 2026

%\cite{Andriot:2018wzk}
\bibitem{Andriot:2018wzk}
D.~Andriot,
%``On the de Sitter swampland criterion,''
Phys. Lett. B \textbf{785}, 570-573 (2018)
%doi:10.1016/j.physletb.2018.09.022
[arXiv:1806.10999 [hep-th]].
%253 citations counted in INSPIRE as of 25 Sep 2026

%\cite{Garg:2018reu}
\bibitem{Garg:2018reu}
S.~K.~Garg and C.~Krishnan,
%``Bounds on Slow Roll and the de Sitter Swampland,''
JHEP \textbf{11}, 075 (2019)
%doi:10.1007/JHEP11(2019)075
[arXiv:1807.05193 [hep-th]].
%750 citations counted in INSPIRE as of 25 Sep 2026

%\cite{Cicoli:2018kdo}
\bibitem{Cicoli:2018kdo}
M.~Cicoli, S.~De Alwis, A.~Maharana, F.~Muia and F.~Quevedo,
%``De Sitter vs Quintessence in String Theory,''
Fortsch. Phys. \textbf{67}, no.1-2, 1800079 (2019)
%doi:10.1002/prop.201800079
[arXiv:1808.08967 [hep-th]].
%239 citations counted in INSPIRE as of 25 Sep 2026

%\cite{Ooguri:2018wrx}
\bibitem{Ooguri:2018wrx}
H.~Ooguri, E.~Palti, G.~Shiu and C.~Vafa,
%``Distance and de Sitter Conjectures on the Swampland,''
Phys. Lett. B \textbf{788}, 180-184 (2019)
%doi:10.1016/j.physletb.2018.11.018
[arXiv:1810.05506 [hep-th]].
%862 citations counted in INSPIRE as of 25 Sep 2026

%\cite{Andriot:2018mav}
\bibitem{Andriot:2018mav}
D.~Andriot and C.~Roupec,
%``Further refining the de Sitter swampland conjecture,''
Fortsch. Phys. \textbf{67}, no.1-2, 1800105 (2019)
%doi:10.1002/prop.201800105
[arXiv:1811.08889 [hep-th]].
%161 citations counted in INSPIRE as of 25 Sep 2026


%\cite{Planck:2018jri}
\bibitem{Planck:2018jri}
Y.~Akrami \textit{et al.} [Planck],
%``Planck 2018 results. X. Constraints on inflation,''
Astron. Astrophys. \textbf{641}, A10 (2020)
%doi:10.1051/0004-6361/201833887
[arXiv:1807.06211 [astro-ph.CO]].
%4745 citations counted in INSPIRE as of 16 Sep 2026

%\cite{Lewandowski:2013aka}
\bibitem{Lewandowski:2013aka}
M.~Lewandowski and A.~Perko,
%``Leading slow roll corrections to the volume of the universe and the entropy bound,''
JHEP \textbf{12}, 060 (2014)
%doi:10.1007/JHEP12(2014)060
[arXiv:1309.6705 [hep-th]].
%13 citations counted in INSPIRE as of 24 Sep 2026

%\cite{Cotler:2026lna}
\bibitem{Cotler:2026lna}
J.~Cotler, V.~Ivo and J.~Maldacena,
%``Renormalization in Liouville gravity and stochastic inflation,''
[arXiv:2608.26244 [hep-th]].
%0 citations counted in INSPIRE as of 17 Sep 2026

%\cite{Cotler:2026flq}
\bibitem{Cotler:2026flq}
J.~Cotler, V.~Ivo and J.~Maldacena,
%``Logarithmic gravity from a very slow roll limit of inflation,''
[arXiv:2609.38052 [hep-th]].
%0 citations counted in INSPIRE as of 03 Oct 2026

%\cite{Polyakov:1981rd}
\bibitem{Polyakov:1981rd}
A.~M.~Polyakov,
%``Quantum Geometry of Bosonic Strings,''
Phys. Lett. B \textbf{103}, 207-210 (1981)
%doi:10.1016/0370-2693(81)90743-7
%3306 citations counted in INSPIRE as of 24 Sep 2026

%\cite{Knizhnik:1988ak}
\bibitem{Knizhnik:1988ak}
V.~G.~Knizhnik, A.~M.~Polyakov and A.~B.~Zamolodchikov,
%``Fractal Structure of 2D Quantum Gravity,''
Mod. Phys. Lett. A \textbf{3}, 819 (1988)
%doi:10.1142/S0217732388000982
%1511 citations counted in INSPIRE as of 24 Sep 2026



%\cite{Mukhanov:1985rz}
\bibitem{Mukhanov:1985rz}
V.~F.~Mukhanov,
%``Gravitational Instability of the Universe Filled with a Scalar Field,''
JETP Lett. \textbf{41}, 493-496 (1985)
%561 citations counted in INSPIRE as of 16 Sep 2026

%\cite{Sasaki:1986hm}
\bibitem{Sasaki:1986hm}
M.~Sasaki,
%``Large Scale Quantum Fluctuations in the Inflationary Universe,''
Prog. Theor. Phys. \textbf{76}, 1036 (1986)
%doi:10.1143/PTP.76.1036
%783 citations counted in INSPIRE as of 16 Sep 2026

%\cite{Seo:2025tsw}
\bibitem{Seo:2025tsw}
M.~S.~Seo,
%``Local gauge invariant operator on isometry breaking background,''
JHEP \textbf{03}, 023 (2026)
%doi:10.1007/JHEP03(2026)023
[arXiv:2510.22879 [hep-th]].
%2 citations counted in INSPIRE as of 25 Sep 2026

%\cite{Seo:2026kxs}
\bibitem{Seo:2026kxs}
M.~S.~Seo,
%``Implication of dressed form of relational observable on von Neumann algebra,''
Class. Quant. Grav. \textbf{43}, no.18, 185010 (2026)
%doi:10.1088/1361-6382/aea5c8
[arXiv:2603.25990 [hep-th]].
%0 citations counted in INSPIRE as of 25 Sep 2026



%\cite{Mukhanov:1981xt}
\bibitem{Mukhanov:1981xt}
V.~F.~Mukhanov and G.~V.~Chibisov,
%``Quantum Fluctuations and a Nonsingular Universe,''
JETP Lett. \textbf{33}, 532-535 (1981)
%2508 citations counted in INSPIRE as of 23 Sep 2026

%\cite{Starobinsky:1982ee}
\bibitem{Starobinsky:1982ee}
A.~A.~Starobinsky,
%``Dynamics of Phase Transition in the New Inflationary Universe Scenario and Generation of Perturbations,''
Phys. Lett. B \textbf{117}, 175-178 (1982)
%doi:10.1016/0370-2693(82)90541-X
%3154 citations counted in INSPIRE as of 23 Sep 2026

%\cite{Guth:1982ec}
\bibitem{Guth:1982ec}
A.~H.~Guth and S.~Y.~Pi,
%``Fluctuations in the New Inflationary Universe,''
Phys. Rev. Lett. \textbf{49}, 1110-1113 (1982)
%doi:10.1103/PhysRevLett.49.1110
%2782 citations counted in INSPIRE as of 23 Sep 2026

%\cite{Hawking:1982cz}
\bibitem{Hawking:1982cz}
S.~W.~Hawking,
%``The Development of Irregularities in a Single Bubble Inflationary Universe,''
Phys. Lett. B \textbf{115}, 295 (1982)
%doi:10.1016/0370-2693(82)90373-2
%2347 citations counted in INSPIRE as of 23 Sep 2026


%\cite{Lidsey:1995np}
\bibitem{Lidsey:1995np}
J.~E.~Lidsey, A.~R.~Liddle, E.~W.~Kolb, E.~J.~Copeland, T.~Barreiro and M.~Abney,
%``Reconstructing the inflation potential : An overview,''
Rev. Mod. Phys. \textbf{69}, 373-410 (1997)
%doi:10.1103/RevModPhys.69.373
[arXiv:astro-ph/9508078 [astro-ph]].
%912 citations counted in INSPIRE as of 16 Sep 2026


%\cite{Lyth:1998xn}
\bibitem{Lyth:1998xn}
D.~H.~Lyth and A.~Riotto,
%``Particle physics models of inflation and the cosmological density perturbation,''
Phys. Rept. \textbf{314}, 1-146 (1999)
%doi:10.1016/S0370-1573(98)00128-8
[arXiv:hep-ph/9807278 [hep-ph]].
%2332 citations counted in INSPIRE as of 16 Sep 2026

%\cite{Riotto:2002yw}
\bibitem{Riotto:2002yw}
A.~Riotto,
%``Inflation and the theory of cosmological perturbations,''
ICTP Lect. Notes Ser. \textbf{14}, 317-413 (2003)
[arXiv:hep-ph/0210162 [hep-ph]].
%632 citations counted in INSPIRE as of 16 Sep 2026

%\cite{Liddle:1992wi}
\bibitem{Liddle:1992wi}
A.~R.~Liddle and D.~H.~Lyth,
%``COBE, gravitational waves, inflation and extended inflation,''
Phys. Lett. B \textbf{291}, 391-398 (1992)
%doi:10.1016/0370-2693(92)91393-N
[arXiv:astro-ph/9208007 [astro-ph]].
%497 citations counted in INSPIRE as of 16 Sep 2026

%\cite{Stewart:1993bc}
\bibitem{Stewart:1993bc}
E.~D.~Stewart and D.~H.~Lyth,
%``A More accurate analytic calculation of the spectrum of cosmological perturbations produced during inflation,''
Phys. Lett. B \textbf{302}, 171-175 (1993)
%doi:10.1016/0370-2693(93)90379-V
[arXiv:gr-qc/9302019 [gr-qc]].
%697 citations counted in INSPIRE as of 16 Sep 2026

%\cite{Kosowsky:1995aa}
\bibitem{Kosowsky:1995aa}
A.~Kosowsky and M.~S.~Turner,
%``CBR anisotropy and the running of the scalar spectral index,''
Phys. Rev. D \textbf{52}, R1739-R1743 (1995)
%doi:10.1103/PhysRevD.52.R1739
[arXiv:astro-ph/9504071 [astro-ph]].
%311 citations counted in INSPIRE as of 16 Sep 2026

%\cite{Schwarz:2001vv}
\bibitem{Schwarz:2001vv}
D.~J.~Schwarz, C.~A.~Terrero-Escalante and A.~A.~Garcia,
%``Higher order corrections to primordial spectra from cosmological inflation,''
Phys. Lett. B \textbf{517}, 243-249 (2001)
%doi:10.1016/S0370-2693(01)01036-X
[arXiv:astro-ph/0106020 [astro-ph]].
%301 citations counted in INSPIRE as of 24 Sep 2026 


%\cite{Redner:2001}
\bibitem{Redner:2001}
S.~Redner, A guide to first-passage processes (Cambridge, 2001)

 



\end{thebibliography}
\end{document}